\documentclass[cameraready]{Interspeech}
\usepackage{multirow}
\usepackage{colortbl}
\usepackage{cleveref}
\usepackage{etoolbox}
\patchcmd{\maketitle}{indicates the corresponding author.}{indicates the corresponding authors.}{}{\PackageError{LARA-Gen}{Corresponding-author patch failed}{Check the class definition.}}
\patchcmd{\thebibliography}{\section{References}}{\section{references}}{}{\PackageError{LARA-Gen}{References-heading patch failed}{Check the class definition.}}

\title{LARA-Gen: Enabling Continuous Emotion Control for Music Generation Models via Latent Affective Representation Alignment}

\author[affiliation={1},equalcontribution]{Jiahao}{Mei}
\author[affiliation={1},equalcontribution]{Shimeng}{Li}
\author[affiliation={2}]{Xuenan}{Xu}
\author[affiliation={1}]{Leyuan}{Jin}
\author[affiliation={3}]{Yixuan}{Yuan}
\renewcommand{\authorsep}{, \protect\\}
\author[affiliation={1}]{Zeyu}{Xie}
\author[affiliation={1}]{Zihao}{Zheng}
\author[affiliation={1}]{Ye}{Tao}
\author[affiliation={3},correspondingauthor]{Yue}{Ding}
\author[affiliation={1},correspondingauthor]{Mengyue}{Wu}

\address{
    $^1$ X-LANCE Lab, Shanghai Jiao Tong University, China \\
    $^2$ Shanghai AI Lab, China \\
    $^3$ Shanghai Mental Health Center, Shanghai Jiao Tong University School of Medicine, China
}

\email{} 


\usepackage{comment}

\begin{document}

\maketitle

\begin{abstract}
Recent text-to-music models enable coherent music generation from text prompts, yet fine-grained emotional control remains unresolved. We introduce \textbf{LARA-Gen}, a framework for continuous emotion control that aligns the model's internal hidden states with representations from an external music understanding model through \textbf{L}atent \textbf{A}ffective \textbf{R}epresentation \textbf{A}lignment~(LARA), enabling effective training. 
In addition, we design an emotion control module based on a continuous valence–arousal space, disentangling emotional attributes from textual content and bypassing the bottlenecks of text-based prompting.
Furthermore, we establish a benchmark with a held-out test set and a robust Emotion Predictor for objective evaluation of emotional controllability.
Experiments demonstrate that LARA-Gen achieves continuous, fine-grained emotion control and a favorable balance between emotion adherence and music quality.
Generated samples are available at \href{https://anonymous2232330.github.io/laragen-web/}{\textcolor{cyan}{\textit{https://anonymous2232330.github.io/laragen-web/}}}.
\end{abstract}

\begin{figure*}[ht]
    \centering
    \includegraphics[width=1\linewidth]{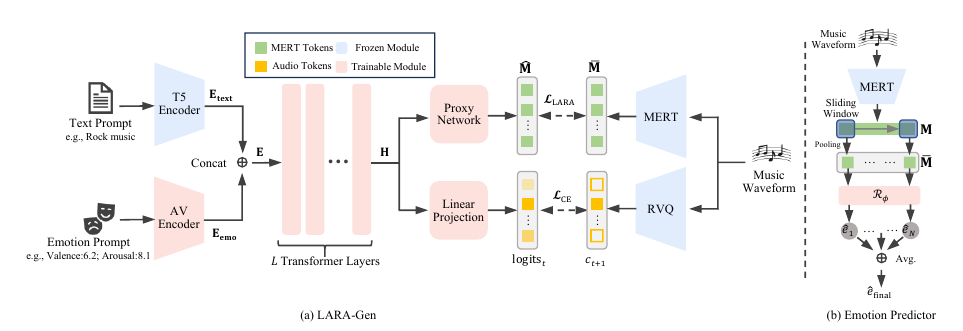}
    \caption{
    \textbf{(a)} LARA-Gen framework. A Proxy Network $\mathcal{P}_{\theta}$ aligns the internal hidden states $\mathbf{H}$ of the backbone model with target features $\bar{\mathbf{M}}$ from a frozen MERT encoder.
    \textbf{(b)} The architecture of the Emotion Predictor. It uses a sliding window over MERT features and an Emotion Regression Head $\mathcal{R}_{\phi}$ to produce a final valence-arousal prediction from an input music clip.
}
    \label{fig:arch}
\end{figure*}

\section{introduction}

Recent advances in text-to-music generation have produced models capable of creating coherent music from textual prompts~\cite{musiclm,mustango,musicgen,liu2023audioldm}. However, achieving fine-grained control over the generated output remains a significant challenge. While some research has begun to explore controllable generation using musical attributes such as melody, rhythm, or structure~\cite{musiccontrolnet,teadapter,tvc}, these efforts have largely overlooked the critical challenge of precise emotional regulation.
A fundamental limitation of existing systems is their reliance on textual descriptions for emotion conditioning (e.g., ``happy'', ``sad''), which suffer from inherent semantic ambiguity. Such descriptors often fail to capture subtle distinctions between emotions (e.g., ``melancholic'' vs. ``sorrowful'') and struggle with rare or complex emotional concepts. More importantly, \textbf{current models lack the capability to handle continuous, numerical emotion descriptors}, which are essential for achieving fine-grained and unambiguous control. This prevents the use of well-established psychological frameworks such as the valence-arousal model~\cite{russell}, despite its ability to represent emotional states in a continuous and interpretable manner. 

The ability to accurately control musical emotion holds significant promise for both general and specialized applications. As a universally perceived quality, emotion represents an intuitive control signal that can make music generation more accessible to non-experts. Furthermore, fine-grained emotional controllability could enable new applications in areas such as music therapy~\cite{de2022music,dash2024ai}, where affective disorders pose a major public health challenge~\cite{huang2019prevalence}, as well as in interactive media and affective computing. However, effectively deploying generative systems in these domains requires overcoming three key challenges: (1) Absence of robust objective metrics for quantifying emotional controllability. Existing objective metrics for music generation (e.g., FAD~\cite{fad} or CLAP~\cite{clap}) primarily assess audio quality or the semantic alignment between a prompt and its generated content, failing to quantify a model's ability to accurately adhere to an emotional target; (2) Inherent ambiguity of textual emotion prompting and the inability of models to process fine-grained emotional attributes; and (3) Inefficiency of implicit training paradigms in capturing subtle emotional characteristics. Conventional autoregressive language model training relies solely on the cross-entropy loss over acoustic tokens. Such indirect and implicit supervision is inefficient and suboptimal for learning the complex mapping from low-dimensional emotion conditions to high-dimensional acoustic features, as subtle emotional characteristics are difficult to capture without explicit supervision~\cite{carbonell2021comparing}.

Inspired by \textbf{Rep}resentation \textbf{A}lignment (REPA)~\cite{repa} in the visual domain, we introduce \textbf{LARA-Gen}, a novel framework that supervises the training process via \textbf{L}atent \textbf{A}ffective \textbf{R}epresentation \textbf{A}lignment~(LARA). By aligning the model's internal representations with rich features from an audio understanding model (MERT~\cite{mert}), LARA-Gen effectively learns the complex mapping from continuous emotion conditions to musical outputs.
Across our experiments, LARA-Gen combines precise control over continuous valence-arousal targets with a favorable balance between audio quality and diversity. To the best of our knowledge, LARA-Gen is among the first approaches to introduce continuous numerical valence-arousal conditioning into pretrained waveform-level text-to-music generation models, representing a paradigm shift from ambiguous textual conditioning to precise affective control. Our key contributions are as follows:
\begin{itemize}

\item  We propose a novel conditioning mechanism that enables generative models to accept continuous valence-arousal values as input, effectively decoupling emotional attributes from textual content and bypassing the limitations of text-based emotion prompting.

\item  We introduce LARA-Gen, a novel generation framework that leverages  \textbf{L}atent \textbf{A}ffective \textbf{R}epresentation \textbf{A}lignment to provide explicit supervision and overcome the inefficiency of standard cross-entropy training.



\item  We introduce a robust Emotion Predictor that enables standardized, objective assessment of emotional controllability by comparing predicted and target VA values.
\end{itemize}

\section{method}
\label{sec:method}

Figure~\ref{fig:arch} illustrates the overall architecture of LARA-Gen (left) and the proposed Emotion Predictor (right).
\subsection{Latent Affective Representation Alignment}
Our framework is built upon a Transformer-based language model, $\mathcal{F}_{\text{LM}}$, which serves as the generative backbone. To enable decoupled emotion control, we process two types of prompts: a text prompt, $\text{p}_{\text{text}}$, for musical content, and a continuous emotion tuple, $\text{p}_{\text{emo}} = (v, a)$, for affective conditioning, where $v, a \in [1, 9]$ are valence and arousal values. 
These prompts are encoded into embeddings using the T5 encoder~\cite{t5} and the Arousal-Valence Encoder~($\text{Encoder}_{\text{AV}}$), respectively: $\mathbf{E}_{\text{text}} = \text{Encoder}_{\text{T5}}(\text{p}_{\text{text}})$ and $\mathbf{E}_{\text{emo}} = \text{Encoder}_{\text{AV}}(v, a)$. Here, $\mathbf{E}_{\text{text}} \in \mathbb{R}^{B \times T_{\text{T5}} \times D_{\text{T5}}}$, $\mathbf{E}_{\text{emo}} \in \mathbb{R}^{B \times D_{\text{T5}}}$, and $B$, $T$, and $D$ denote the batch size, sequence length, and embedding dimension, respectively. $\text{Encoder}_{\text{AV}}$ is a lightweight Multi-Layer Perceptron (MLP). It maps a two-dimensional tensor representing valence and arousal values (normalized to the range $[-1, 1]$) to a $D_{\text{T5}}$-dimensional vector. The emotion embedding is reshaped to $B \times 1 \times D_{\text{T5}}$ and prepended to the text embeddings to form the final conditioning embedding $\mathbf{E} = \text{Concat}(\mathbf{E}_{\text{emo}}, \mathbf{E}_{\text{text}})$. This combined embedding $\mathbf{E}\in \mathbb{R}^{B \times (T_{\text{T5}}+1) \times D_{\text{T5}}}$ is fed into the cross-attention layers of the backbone model at each Transformer block.

The training objective for LARA-Gen is a composite loss function designed to simultaneously ensure acoustic fidelity and emotional accuracy. To formulate this, we first define the ground-truth representations from a given mono audio waveform $\mathbf{A} \in \mathbb{R}^{B \times T_{\text{wav}}}$. The target for the standard autoregressive task is a sequence of discrete acoustic tokens created by a pretrained residual vector quantization~(RVQ) compression model~\cite{encodec}, $\mathbf{C} = \text{RVQ}(\mathbf{A})$, where $\mathbf{C} \in \mathbb{Z}^{B \times K \times T}$, $K$ is the number of codebooks, $T$ is the sequence length. The target sequences for our novel emotional alignment task are continuous features extracted from the external pretrained audio understanding model MERT~\cite{mert}, $\bar{\mathbf{M}}=\{\bar{\mathbf{m}}_1, \bar{\mathbf{m}}_2, \dots, \bar{\mathbf{m}}_N\}$, each $\bar{\mathbf{m}}_i \in \mathbb{R}^{B \times D_{\text{MERT}}}$, with details provided in Section~\ref{ssec:emotion_predictor}.

The first component of our training objective is Cross-Entropy Loss, $\mathcal{L}_{\text{CE}}$. Let the ground-truth sequence of discrete acoustic tokens be denoted as $\mathbf{C} = (c_1, c_2, \dots, c_T)$, where $c_t$ is the token at timestep $t$. Following the standard teacher-forcing paradigm, we define the model's input sequence as $\mathbf{C}_{\text{in}} = (c_1, \dots, c_{T-1})$ and the corresponding target sequence as $\mathbf{C}_{\text{target}} = (c_2, \dots, c_T)$.
During the forward pass, the backbone model $\mathcal{F}_{\text{LM}}$ processes the input sequence $\mathbf{C}_{\text{in}}$ and the conditioning embedding $\mathbf{E}$ to produce a sequence of hidden states $\mathbf{H}^{(L)} \in \mathbb{R}^{B \times (T-1) \times D}$ at its final layer $L$. These hidden states are then projected through a linear layer to produce a sequence of logit vectors, $\text{Logits} = (\text{logits}_1, \dots, \text{logits}_{T-1})$. The cross-entropy loss is then computed over the entire sequence by comparing the predicted logits at each timestep with the corresponding ground-truth target token:
\begin{equation}
    \mathcal{L}_{\text{CE}} = \mathbb{E}_{(\mathbf{C}, \mathbf{E}) \sim \mathcal{D}} \left[ \sum_{t=1}^{T-1} \text{CrossEntropy}(\text{logits}_t, c_{t+1}) \right]
\end{equation}
where $\mathcal{D}$ represents the data distribution.

The core of our contribution is the Latent Affective Representation Alignment (LARA) Loss, $\mathcal{L}_{\text{LARA}}$. To compute this, we must bridge the gap between the backbone's high-resolution hidden state sequence, $\mathbf{H} \in \mathbb{R}^{B \times T \times D}$, and the lower-resolution target MERT feature tokens, $\bar{\mathbf{M}} \in \mathbb{R}^{B \times N \times D_{\text{MERT}}}$, where $T \gg N$.
We achieve this temporal downsampling with a lightweight, trainable \textbf{Proxy Network}, $\mathcal{P}_{\theta}$, implemented as a Transformer decoder. The network uses a set of $N$ learnable query tokens, $\mathbf{Q}\in \mathbb{R}^{N \times D}$, to summarize the information from the entire hidden state sequence $\mathbf{H}$ (acting as memory) via cross-attention. The updated query sequence is then linearly projected to predict the MERT features, $\hat{\mathbf{M}}$:
\begin{equation}
    \hat{\mathbf{M}} = \text{Linear}(\text{TransformerDecoder}(\mathbf{Q}, \mathbf{H}))
\end{equation}
where $\hat{\mathbf{M}} \in \mathbb{R}^{B \times N \times D_{\text{MERT}}}$. This architecture effectively learns to distill the long sequence of generative representations into a compact sequence of emotion features for alignment.

The LARA loss then minimizes the Mean Squared Error (MSE) between these predicted features $\hat{\mathbf{M}}$ and the ground-truth MERT features $\bar{\mathbf{M}}$:
\begin{equation}
    \mathcal{L}_{\text{LARA}} = \text{MSE}(\hat{\mathbf{M}}, \bar{\mathbf{M}})
\end{equation}
Finally, the total training objective $\mathcal{L}_{\text{total}}$ is a weighted sum of these two losses:
\begin{equation}
    \mathcal{L}_{\text{total}} = \mathcal{L}_{\text{CE}} + \alpha \cdot \mathcal{L}_{\text{LARA}}
\end{equation}
where $\alpha$ is a hyperparameter that balances the two objectives. 
By optimizing this composite loss, LARA-Gen generates high-quality music that is acoustically faithful and emotionally precise.

\subsection{Emotion Predictor for Objective Evaluation}
\label{ssec:emotion_predictor}

To establish a reproducible emotional music generation benchmark, we introduce an Emotion Predictor, $\mathcal{E}_{\phi}$, which provides a quantitative metric for emotional accuracy. Trained on multiple public music emotion datasets for robustness, it remains frozen as a fixed evaluator.
The predictor consists of a frozen pretrained MERT audio encoder~\cite{mert} and a trainable \textbf{Emotion Regression Head}, $\mathcal{R}_{\phi}$, which learns the non-linear mapping from acoustic features to valence-arousal space.

Let a given audio waveform be $\mathbf{A} \in \mathbb{R}^{B \times T_{\text{wav}}}$. We first extract the target MERT feature sequence $\mathbf{M} = \text{MERT}(\mathbf{A})$, where $\mathbf{M} \in \mathbb{R}^{B \times T_{\text{MERT}} \times D_{\text{MERT}}}$. To robustly capture the emotional content over time, we analyze the feature sequence using a sliding window approach instead of a single global pooling operation.
We define a sliding window of length $W$ seconds, which corresponds to $W_{\text{tokens}}$ timesteps in the MERT feature sequence, with a stride of $S$ seconds ($S = W$). This process segments the full feature sequence $\mathbf{M}$ into $N$ shorter segments, $\{\mathbf{m}_1, \mathbf{m}_2, \dots, \mathbf{m}_N\}$, where each $\mathbf{m}_i \in \mathbb{R}^{B \times W_{\text{tokens}} \times D_{\text{MERT}}}$.

For each segment $\mathbf{m}_i$, we first apply temporal mean pooling to obtain a fixed-size representation $\bar{\mathbf{m}}_i = \text{Pool}(\mathbf{m}_i)$. Each pooled segment is then independently processed by the Emotion Regression Head (an MLP), yielding a sequence of segmental valence-arousal predictions $(\hat{v}_i, \hat{a}_i)$ as $\hat{\mathbf{e}}_i = \mathcal{R}_{\phi}(\bar{\mathbf{m}}_i) \in \mathbb{R}^{B \times 2}$ for $i=1,\dots,N$. Finally, to obtain a single emotion prediction for the entire input audio clip, we compute the overall emotion tuple as the average of all segmental outputs: $\hat{\mathbf{e}}_{\text{final}} = \frac{1}{N}\sum_{i=1}^{N}\hat{\mathbf{e}}_i$.


Our segmental approach ensures the Emotion Predictor captures temporal variations for a stable, representative emotion assessment. 
The Regression Head minimizes the Mean Squared Error (MSE) between the final valence-arousal prediction $\hat{\mathbf{e}}_{\text{final}}$ and the ground-truth annotation $\mathbf{e}$.

\begin{figure*}[t]
    \centering
    \includegraphics[width=0.95\linewidth]{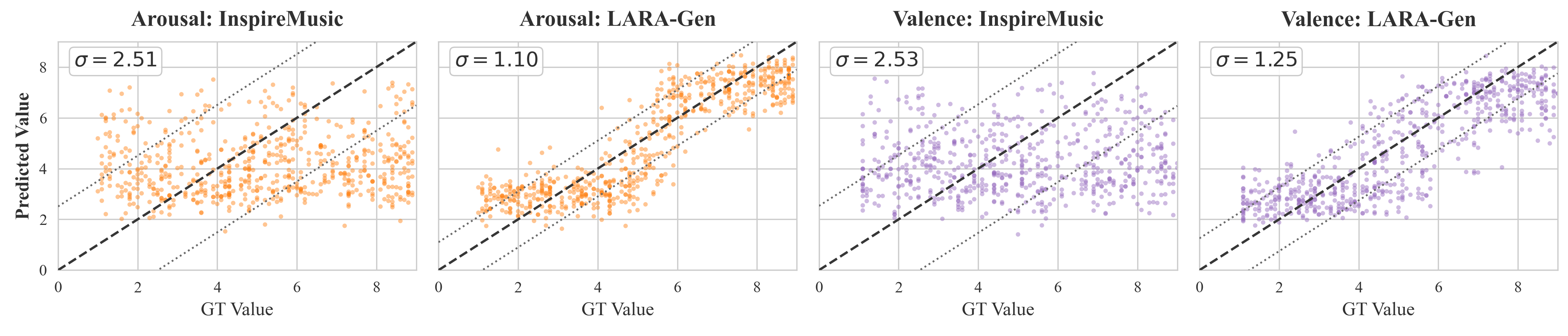}
    \caption{Emotion values predicted by our Emotion Predictor vs. ground-truth emotion values on the MUSE-VA test set. The LARA-Gen system outperforms the InspireMusic baseline in both error and correlation.}
    \label{fig:scatter}
\end{figure*}

\begin{table*}[ht]
\caption{Emotional music generation performance on the MUSE-VA test set. Emotion consistency is evaluated by our proposed Emotion Predictor. Best and second-best results among generated systems are \textbf{bold} and \underline{underlined}, respectively.}
\label{tab:main}
\centering
\small
\setlength{\tabcolsep}{3pt}
\renewcommand{\arraystretch}{1.1}
\begin{tabular*}{\textwidth}{@{\extracolsep{\fill}}lcc*{8}{c}@{}}
\toprule
\multirow{2}{*}{Generation System} & \multirow{2}{*}{Prompt} & \multirow{2}{*}{Loss} & \multirow{2}{*}{FAD$\downarrow$} & \multirow{2}{*}{Pairwise L2$\uparrow$} & \multicolumn{3}{c}{Arousal} & \multicolumn{3}{c}{Valence} \\
\cmidrule(lr){6-8}\cmidrule(lr){9-11}
 & & & & & CCC$\uparrow$ & PCC$\uparrow$ & RMSE$\downarrow$ & CCC$\uparrow$ & PCC$\uparrow$ & RMSE$\downarrow$ \\
\midrule
Ground Truth & - & - & 0.000 & 2.963 & 0.868 & 0.875 & 0.283 & 0.829 & 0.842 & 0.309 \\
MusicGen & Text & - & 3.060 & 2.743 & 0.092 & 0.153 & 0.676 & 0.129 & 0.190 & 0.623 \\
InspireMusic & Text & - & 3.325 & \underline{3.454} & 0.060 & 0.083 & 0.673 & 0.044 & 0.058 & 0.659 \\
ACE-Step & Text & - & 5.198 & 1.858 & 0.088 & 0.214 & 0.622 & 0.064 & 0.133 & 0.656 \\
Emotion Text Fine-tuning & Text & CE & \underline{2.153} & 2.947 & 0.795 & 0.816 & 0.340 & 0.700 & 0.763 & 0.401 \\
LARA-Gen & VA & CE + LARA & \textbf{1.888} & 2.995 & \underline{0.869} & \textbf{0.878} & \textbf{0.279} & \textbf{0.813} & \textbf{0.837} & \textbf{0.322} \\
\quad w/o LARA & VA & CE & 3.645 & \textbf{3.602} & \textbf{0.873} & \underline{0.875} & \underline{0.280} & \underline{0.776} & \underline{0.802} & \underline{0.350} \\
\bottomrule
\end{tabular*}

\end{table*}

\begin{table}[ht]
\centering
\caption{Subjective evaluation results. Significance levels for PCC are denoted by asterisks ($^{*} p < 0.05$, $^{***} p < 0.001$). Best results among generated systems are highlighted in \textbf{bold}.}
\label{tab:subjective_evaluation}
\begin{tabular}{lccc}
\toprule
\multirow{2}{*}{Generation System} & \multirow{2}{*}{OVL $\uparrow$} & \multicolumn{2}{c}{PCC$\uparrow$} \\
\cmidrule(lr){3-4}
 & & Arousal & Valence \\
\midrule
Ground Truth & 3.90{\scriptsize $\pm$0.97} & 0.882$^{***}$ & 0.549$^{*}$ \\
\midrule
Text Fine-tuning & 3.16{\scriptsize $\pm$1.07} & 0.709$^{*}$ & 0.613$^{*}$ \\
LARA-Gen & \textbf{3.47{\scriptsize $\pm$1.14}} & \textbf{0.881}$^{***}$ & \textbf{0.648}$^{*}$ \\
\bottomrule
\end{tabular}
\par\vspace{-10pt}
\end{table}

\section{experiments}
\label{sec:experiments}

\noindent\textbf{Datasets.}
\label{ssec:datasets}
Music generation experiments are conducted on MUSE-VA~\cite{museva}, a multimodal corpus comprising 6,254 synthetic music samples annotated in the continuous valence-arousal space~(range 1--9). We adopt its predefined training, validation, and test partitions of 5,004, 625, and 625 samples, respectively. In order to reduce dataset-specific bias in the Emotion Predictor, we train it jointly on the training splits of MUSE-VA, DEAM~\cite{mead}, and PMEmo~\cite{pmemo}, mapping their annotations to a common $[-1,1]$ scale.

We evaluate emotional controllability on 625 held-out MUSE-VA recordings, using their valence-arousal annotations as generation targets and initial 30-second segments as reference audio. After standardization to 32\,kHz mono, generated and reference audio are assessed by the frozen Emotion Predictor against the corresponding valence-arousal annotations. These test recordings are excluded from generator fine-tuning and predictor training.

\noindent\textbf{Model Specifications and Baselines.} We use the pretrained MusicGen-Medium~\cite{musicgen} as our generative backbone, retaining its original T5 text encoder and RVQ compression settings. Our \textbf{Emotion Predictor} consists of a frozen MERT-95M~\cite{mert} backbone that extracts features using 5-second non-overlapping windows. The regression head is implemented as a three-layer MLP trained to optimize the MSE loss, utilizing an initial learning rate of $10^{-4}$ and a weight decay of $10^{-4}$. To eliminate the influence of text-side semantic variations and isolate the emotional control capability, the text prompt is fixed to ``Generate an instrumental music with arousal $a$ and valence $v$.'' We compare our approach against the \textbf{Emotion Text Prompting baselines} under two settings: zero-shot inference with pretrained models, and Emotion Text Fine-tuning on our dataset using cross-entropy loss. Both settings map each target valence-arousal coordinate to the nearest emotion word in the ANEW~\cite{anew} lexicon, yielding the prompt ``generate a [emotion word] music''. The zero-shot baselines comprise MusicGen-Medium, InspireMusic-1.5B~\cite{inspiremusic}, and ACE-Step~\cite{acestep15}, while Emotion Text Fine-tuning uses MusicGen-Medium. All models undergoing fine-tuning were trained for 37,000 steps, and for LARA-Gen, the loss weight $\alpha$ was set to 100.

\noindent\textbf{Evaluation Metrics.}
\label{ssec:metrics}
For objective evaluation, we assess emotion control accuracy using the Concordance Correlation Coefficient (CCC), Pearson Correlation Coefficient (PCC), and Root Mean Square Error (RMSE) between the target valence-arousal values and the predictions extracted from generated samples using our proposed Emotion Predictor. Music quality is measured using the Fréchet Audio Distance (FAD) metric, while diversity is quantified by the mean pairwise Euclidean distance between clip-level VGGish embeddings (Pairwise L2). For the subjective evaluation, we recruited 10 participants (all non-music-major university students). Each participant rated 48 music clips. Participants annotated the Overall Music Quality (OVL, range 1-5) and perceived Valence-Arousal values (range 1-9). To robustly measure subjective emotion adherence, we calculated the PCC and its statistical significance ($p$-value) between human ratings and the target labels. For the 10 raters selected by model performance, exploratory reliability analysis of the 48 clips in Table~\ref{tab:subjective_evaluation} used the absolute-agreement ICC(2,10), yielding 0.749 for OVL, 0.965 for arousal, and 0.810 for valence. These estimates describe the reliability of ratings averaged across the 10 participants.

\section{results}
\label{sec:results}

We evaluate the performance of our proposed LARA-Gen framework against baselines and ground truth. We first analyze the performance of our Emotion Predictor to establish a reliable evaluation baseline, and then present a comprehensive analysis of the music generation systems using both objective and subjective metrics. The results are shown in ~\Cref{tab:main} and \Cref{tab:subjective_evaluation}.


\noindent\textbf{Emotion Predictor Performance.}
We validate the proposed Emotion Predictor on reference audio from the held-out MUSE-VA test set (\Cref{tab:main}, GT row). The frozen predictor achieves strong agreement with the reference annotations ($CCC_A=0.868$, $CCC_V=0.829$). The high correlations and low prediction errors across both dimensions support its use as a reliable evaluator of emotional consistency in music generation.


\noindent\textbf{Generation Quality.}
All three zero-shot text-prompting baselines yield higher FAD scores than the fine-tuned text baseline and LARA-Gen, highlighting limitations in matching the target audio distribution without task-specific adaptation. With continuous VA conditioning and MERT feature alignment, LARA-Gen achieves a balanced combination of fidelity (FAD=1.888) and diversity (Pairwise L2=2.995). LARA-Gen without LARA achieves greater diversity (Pairwise L2=3.602) without alignment constraints but incurs a higher FAD (3.645). The subjective OVL results show that LARA-Gen receives a higher mean rating than Text Fine-tuning and approaches the rating of ground truth.

\noindent\textbf{Emotion Control Accuracy.}
Objective metrics demonstrate that continuous VA prompts substantially enhance emotion control, with LARA-Gen and the ablation without LARA occupying the top two positions among generated systems across all emotion metrics. For arousal, LARA-Gen reaches the highest PCC (0.878) and lowest RMSE (0.279), while its CCC (0.869) is slightly below that of LARA-Gen without LARA (0.873). For valence, LARA-Gen (CCC=0.813, PCC=0.837) outperforms LARA-Gen without LARA (CCC=0.776, PCC=0.802) and approaches the performance of GT (CCC=0.829, PCC=0.842). These results suggest that LARA effectively guides music generation toward the target valence-arousal values. Subjectively, LARA-Gen achieves strong arousal correlation and significant valence correlation, with higher PCC values than Text Fine-tuning for both arousal (0.881 vs. 0.709) and valence (0.648 vs. 0.613).


\noindent\textbf{Error and Distribution Analysis.}
As shown in~\Cref{fig:scatter}, LARA-Gen exhibits superior stability. Compared to the text baseline, it forms a tighter distribution around the ideal fit with lower error standard deviations. Consistently, LARA-Gen achieves the lowest arousal RMSE (0.279), ranking first among all evaluated models on this metric. These complementary results support the effectiveness of LARA-Gen in translating continuous affective targets into music with well-defined emotional characteristics.


\section{conclusion}
In this work, we presented \textbf{LARA-Gen}, a novel framework enabling continuous and fine-grained emotional control in music generation models. To overcome the semantic ambiguity of text prompts and the inefficiency of conventional cross-entropy training, we proposed a continuous valence-arousal conditioning mechanism and \textbf{L}atent \textbf{A}ffective \textbf{R}epresentation \textbf{A}lignment (LARA). LARA provides explicit, dense supervision by aligning the generative backbone's hidden states with rich acoustic features from an external audio understanding model, while our robust Emotion Predictor enables objective evaluation of emotional controllability. Experimental results demonstrate that LARA-Gen achieves a favorable balance among emotion control accuracy, generation quality, and diversity. By enabling precise numerical control and providing a standardized objective metric, this work paves the way for future research in controllable affective music generation.

\bibliographystyle{IEEEtran}
\setcounter{section}{5}
\bibliography{mybib}

\end{document}